\documentclass[doublecol]{epl2} 
\pdfoutput=1
\usepackage{subfigure}
\usepackage{amssymb}
\usepackage{amsfonts}
\usepackage{epsfig}
\usepackage{graphicx}
\usepackage{color}

\title{Faceted anomalous scaling in the epitaxial growth of semiconductor films}
\shorttitle{Faceted anomalous scaling in the epitaxial growth} 

\author{F. S. Nascimento\and S. C. Ferreira\thanks{On leave at Departament de F\'{\i}sica i Enginyeria Nuclear, Universitat Polit\'ecnica de Catalunya, Barcelona, Spain.} \and S. O. Ferreira}
\shortauthor{F. S. Nascimento \etal}

\institute{                    
  Departamento de F\'{\i}sica - Universidade Federal de Vi\c{c}osa, 36571-000, Vi\c{c}osa,
Minas Gerais, Brazil
}

\abstract{
We apply the generic dynamical scaling theory (GDST) to the surfaces of CdTe 
polycrystalline films grown in glass substrates.  
The analysed data were obtained with a stylus profiler with an estimated resolution
lateral resolution of $l_c=0.3~\mu$m. 
Both real two-point correlation function and power spectrum analyses 
were done. We found that the GDST applied to the surface power spectra 
foresees faceted morphology  in contrast with the self-affine surface indicated by
the local roughness exponent found via the height-height correlation function.
This inconsistency is explained in terms of convolution effects 
resulting from the finite size of the probe tip used to scan the surfaces. 
High resolution AFM images corroborates the predictions of GDST.}

\begin{document}

\maketitle

{The kinetic roughening of interfaces is an outstanding topic of
nonequilibrium Statistical Physics which has been intensively investigated in
both theoretical \cite{Lopez2005,Ramasco,Nicoli,NicoliJstat,Sasamoto} and
experimental \cite{Cordoba,Takeuchi2010,Lafouresse,Ferreira2006,Huo2001}
frontlines.}
A generic dynamical scaling theory (GDST) for evolving interfaces includes both interface 
fluctuations and power spectra (structure factor) in the real and momentum spaces,
respectively~\cite{Ramasco}. The analysis in the real space can be performed using 
the local height-height correlation function 
\begin{equation}
\label{eq:g}
G(l,t)=\langle\overline{[h(x+l,t)-h(x,t)]^2}\rangle,
\end{equation}
where the overbar means averaging over the surface profile and $\langle\cdots\rangle$ 
the averaging over distinct profiles. GDST gives that $G(l,t) = l^{2\alpha}\Phi(l/t^{1/z})$ 
where $\alpha$ and $z$ are the roughness and dynamical exponents, respectively. 
The scaling function $\Phi$ behaves as 
\begin{equation}
 \Phi(x)\sim \left\lbrace \begin{array}{lll}
                           x^{-2(\alpha-\alpha_{loc})} & \mbox{if} & x\ll 1 \\
                           u^{-2\alpha} & \mbox{if} & x\gg 1
                          \end{array}
 \right..
\end{equation}
The local roughness exponent $\alpha_{loc}$ determines the scaling locally.  
The power spectrum, defined as  
$S(k,t) =  \langle \hat{h}(\mathbf{k},t)\hat{h}(-\mathbf{k},t) \rangle$ 
with $\hat{h}$ being the Fourier transform of the height surface,  scales as 
$S(k,t) = k^{-(2\alpha+d)}\Psi(kt^{1/z})$, where the scaling function $\Psi$ is
\begin{equation}
\label{eq:ansatz_power}
 \Psi(x) \sim \left\lbrace \begin{array}{lll}
                            x^{2\alpha+d} & \mbox{if} & x\ll 1 \\
                            x^{2(\alpha-\alpha_s)} & \mbox{if} & x\gg 1
                            \end{array}
 \right..        
\end{equation}
In this scaling function,  $d$ is surface topological dimension, and $\alpha_s$ is the spectral roughness exponent. 

This generic scaling ansatz has been observed in a large collection of models and  
experiments as can be looked up in~\cite{Cordoba,Lopez2005,Lafouresse} and references 
therein. This scaling ansatz implies a constraint between the exponents that depends 
specially on the spectral roughness exponent~\cite{Ramasco,Lopez2005,Lopez1997}. 
If $\alpha_s<1$ the surface is self-affine with spectral and local roughness 
exponents being equal, $\alpha_{loc}=\alpha_s$. If $\alpha_s>1$ the surface is 
locally smooth with $\alpha_{loc}=1$. Each case is still classified into two 
subclasses. For $\alpha_s<1$, if $\alpha_s=\alpha$ we have the regular 
Family-Vicsek (FV) scaling, otherwise the system has the intrinsically anomalous 
scaling. For $\alpha_s>1$, we have the super-roughening scaling if $\alpha_s=\alpha$ 
and faceted growth scaling otherwise. The constraint between $\alpha_s$ and $\alpha_{loc}$ 
is a beautiful analytical result derived from the scaling ansatz~\cite{Lopez1997} while 
the subclasses are allusive to their physical implications~\cite{Lopez2005}, with 
exception of the FV scaling which is due to the ansatz conceivers~\cite{Family}. 

FV, super-rough, and intrinsic scalings have been reported in several experimental 
works~\cite{Cordoba, Lopez2005, Lafouresse} but the faceted one has only been achieved 
quite recently in the electrodissolution of pure polycrystalline iron~\cite{Cordoba}. 
The surfaces undergo a transition from intrinsic to faceted anomalous scalings as 
dissolution time increases. The surfaces in the faceted regime were highly 
anisotropic and, consequently, the scaling analyses were performed with 
one-dimensional profiles, in a direction orthogonal to the anisotropy, in addition to the analyses 
of the two-dimensional surfaces. The faceted anomalous scaling was evident only 
for the one-dimensional case. This phenomenon was ascribed to the averages over 
all directions performed in $d=2$, which underestimate the scaling exponents due 
to the contributions of unclear faceted morphology in some directions. 

We investigate the anomalous roughening in semiconductor CdTe polycrystalline 
thick films grown on glass substrates. The real space scaling of this system was 
previously investigated~\cite{Ferreira2006,Mata2008}. In the present work, we
identify a crossover in the dynamical scaling exponents not noticed in Ref.~\cite{Ferreira2006}
and perform the power spectrum scaling analysis. 
We find that the GDST applied to the surface power spectra 
foresees faceted growth whereas the correlation function analysis in points out for a 
self-affine surface. This inconsistency is explained in terms of the convolution effects 
resulting from the finite size of the probe tip used to scan the surfaces.    


The CdTe films were deposited on glass substrates covered with a transparent 
conducting oxide (TCO), SnO$_2$:F, resulting a rough initial surface. 
The glass/TCO surface has a rms width of $16$~nm that is much larger 
than $3$~nm observed for the pure glass substrate. 
Although the presence of this layer introduces an undesirable large initial roughness,
it is important from an application viewpoint because it is the front contact layer of solar
cells produced with this material \cite{Wu2004}.

The samples used in the present work were produced 
by hot wall epitaxy at a growth rate of approximately 0.1 nm/s. 
Details of the growth system and sample preparation can be found 
elsewhere~\cite{Ferreira2006,Leal}. It is important to mention that the growth of some samples 
was repeated using the molecular beam epitaxy technique and they exhibited the same 
behaviour. The scaling analysis was done in one-dimensional profiles with 300 $\mu$m of length 
and 4570 pixels acquired with a stylus profiler XP1-AMBIOS. Consequently, the topological (scan) 
dimension is $d=1$ independently of the $2+1$-dimensional surface. At least 20 profiles scanned 
at random directions were used in the averaging. 

Despite of the good vertical sensibility 
(better than 1 nm) and step size scan (better than 100~nm) of 
the profiler  the large tip radius ($\approx~1~\mu$m) limits the lateral 
resolution power due to convolution effects \cite{AFM_RMP2003}. 
As a consequence, the data at short scales correspond to the convolution 
of the film surface and the probe tip geometries.
A criterion accepted by the microscopy community establishes that a probe tip 
can resolve peaks separated up to 20\% of its radius. 
Therefore, one can estimate a lateral resolution $l_c\approx0.3~\mu$m 
for this scan device. This lateral resolution, however, does not hinder
the measurement of long wavelength fluctuations and the global roughness 
of the entire profile can be accurately determined. 
The short scale details of the surface morphology were visualized using an atomic force microscope 
(NTMDT-Ntegra Prima) operating in the semi-contact mode. These AFM images were not used in the 
scaling analyses. The use of profiler scans allows to study large amplitude and wavelength 
fluctuations not achievable with a regular AFM.  The growth time was varied from $t=30$ to 660~min 
and the substrate temperatures $T=150,~200,~250$, and 300~$^\circ$C were analysed. As will be shown, reliable scaling of the power spectrum  were obtained for $T=250$ and 300~$^\circ$C. The CdTe films detach from the glass substrates and crack during 
the cooling process for growth times longer than $660$~min. This technical limitation hindered the 
growth of samples beyond this limit.

The exponents $\alpha_{loc}$ and $\alpha_s$ were obtained directly from 
the scaling behaviours of $G(l,t)\sim l^{2\alpha_{loc}}$ and 
$S(k,t)\sim k^{-(2\alpha_s+1)}$ at $t=660$ min. The global interface width is defined as
\begin{equation}
W(t)=\left\langle\overline{[h(x,t)-\overline{h}(t)]^2}\right\rangle^{1/2}\sim t^\beta,
\end{equation}
where $\beta$ is the growth exponent.
GDST states $\alpha_{loc}=\min(1,\alpha_s)$ and $\alpha=\beta z$~\cite{Ramasco,Lopez1999}. 
Even though the local and spectral exponents are not independent, the local roughness exponent 
was measured for a comparison between the theoretical prediction and experimental observation. 

\begin{figure}[ht]
 \centering
 \includegraphics[width=7.5cm]{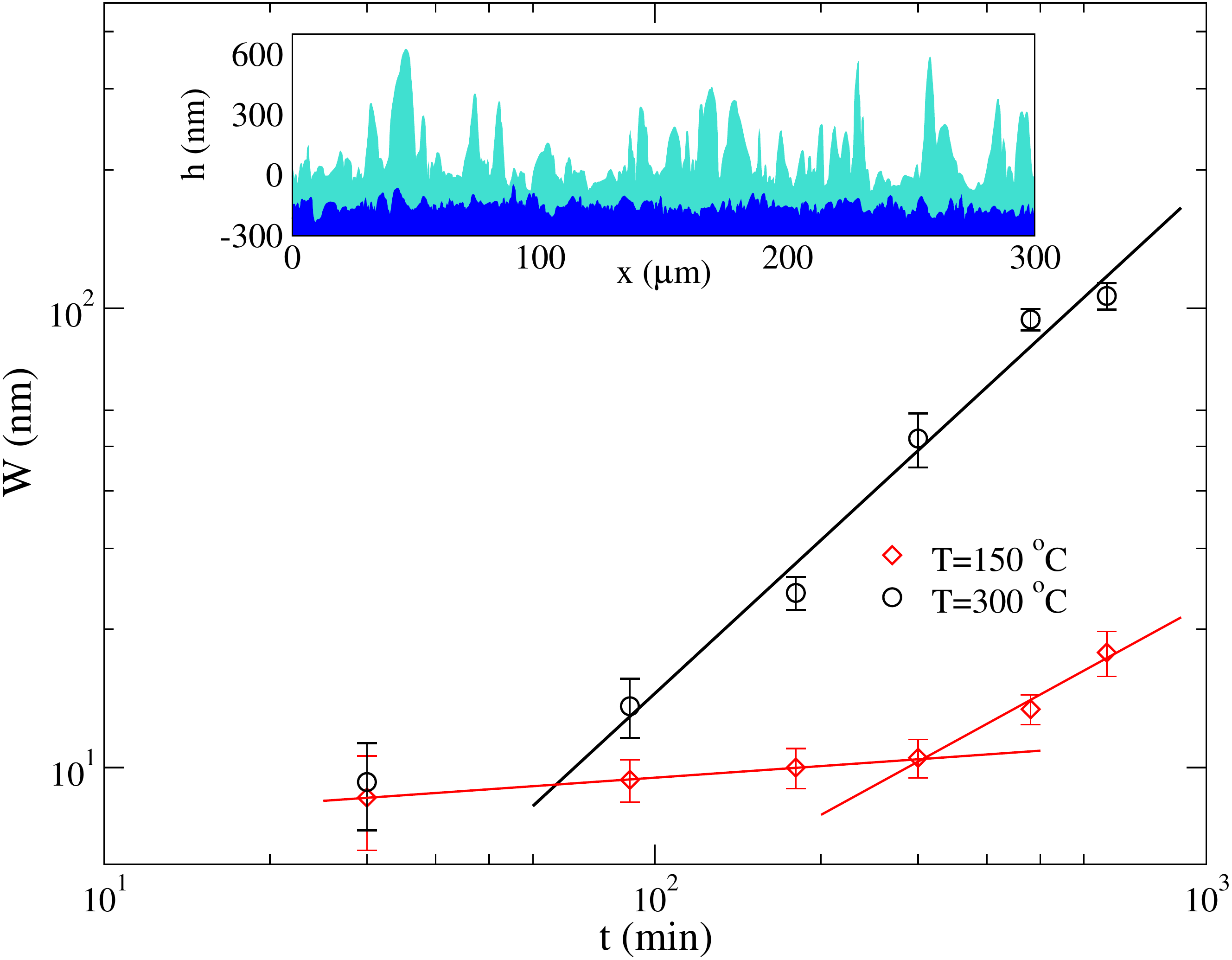}
 \caption{Interface width evolution for two growth temperatures. Lines are power 
law regressions. Inset  shows two surface profiles for growth times $t=180$~min (dark) 
and $t = 660$ min (light) at $T=300~^\circ$C.}
 \label{fig:widths}
\end{figure}

The surface width evolution, which is shown in Fig.~\ref{fig:widths}, has a crossover 
from a regime of low to high growth exponents. This crossover was not perceived in the former 
scaling analysis of this CdTe/TCO/glass system, but it is also present as one can 
check in Ref.~\cite{Ferreira2006}. The crossover is probably on account of the initial surface 
roughness, which initially diminishes due to the occupancy of grooves and 
valleys. The temperature increase enhances the downward funnelling, reducing the crossover times
and, consequently, the largest power-law interval was obtained for $T=300~^\circ$C.
An additional evidence for our proposition 
is that this crossover was not observed in the CdTe deposition on pure glass substrates 
(a smaller initial roughness) at the same experimental conditions~\cite{Leal}. Therefore, 
the growth exponents in Ref.~\cite{Ferreira2006} are under-estimated but the main conclusion 
of that work, stating a $\beta$ exponent depending on temperature, still holds.

Whole scaling analyses were performed for the growth times after 
the crossovers. The $\beta$ exponents are shown in Table~\ref{tab:exp}. The large uncertainties for 
temperatures $150$ and $200~^\circ$C reflect the short time intervals of power law growth. Taking 
into account the uncertainties, these exponents are nearly constant for temperatures 150, 200, and 
250~$^\circ$C. For $T=300~^\circ$C, the growth exponent $\beta>1$ is a signature of an unstable 
growth that is unusually fast in the framework of kinetic roughening. However, self-affine 
(FV scaling) unstable growth was recently found in stochastic 
equations related to nonlocal interface dynamics~\cite{Nicoli,NicoliJstat}. 

Interestingly, the growth exponent obtained numerically for the unstable version
of the stochastic Michelson-Sivashinsky (SMS) equation,
$\beta_{sms}=1.14$~\cite{Nicoli}, coincides with our estimate $\beta=1.11(2)$.
The number in parenthesis represents the uncertainty in last digit meaning
$\beta=1.11\pm0.02$. {The mentioned unstable SMS equation, written in the
momentum space with suitable rescaled parameters, reads as \cite{Nicoli} 
\begin{equation}
 \partial_t\hat{h}= (k-k^2)\hat{h}+\frac{1}{2}\mathcal{F}(|\nabla h|^2)+\hat{\eta},
\end{equation}
where $\mathcal{F}(g)=\hat{g}$ is the Fourier transform of $g$ and $\eta$ a
Gaussian noise. Apart from the term $k\hat{h}$, this equation is the well known
Kardar-Parisi-Zhang equation \cite{KPZ} related to a non-conserved growth. The
instability is due to the term  $k\hat{h}$ that introduces the
non-locality in the equation.} However, as shown in the present work, the
CdTe/Glass system is not self-affine ($\alpha_s\ne\alpha_{loc}$) and cannot be
directly associated with the SMS equation.

\begin{figure}[ht]
 \centering
 \subfigure[\label{fig:power300}]{\includegraphics[width=7.5cm]{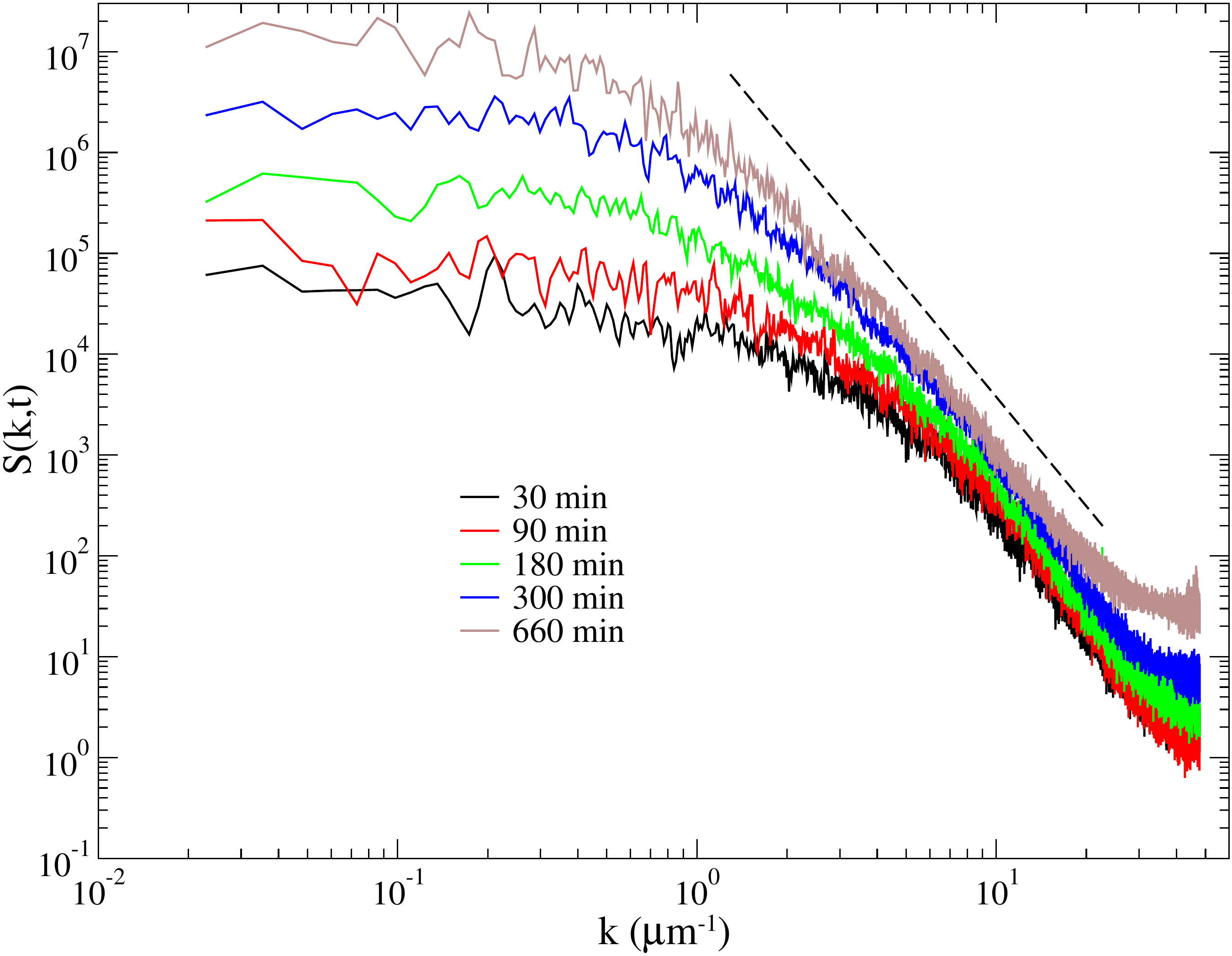}}
 \subfigure[\label{fig:colapses}]{\includegraphics[width=7.5cm]{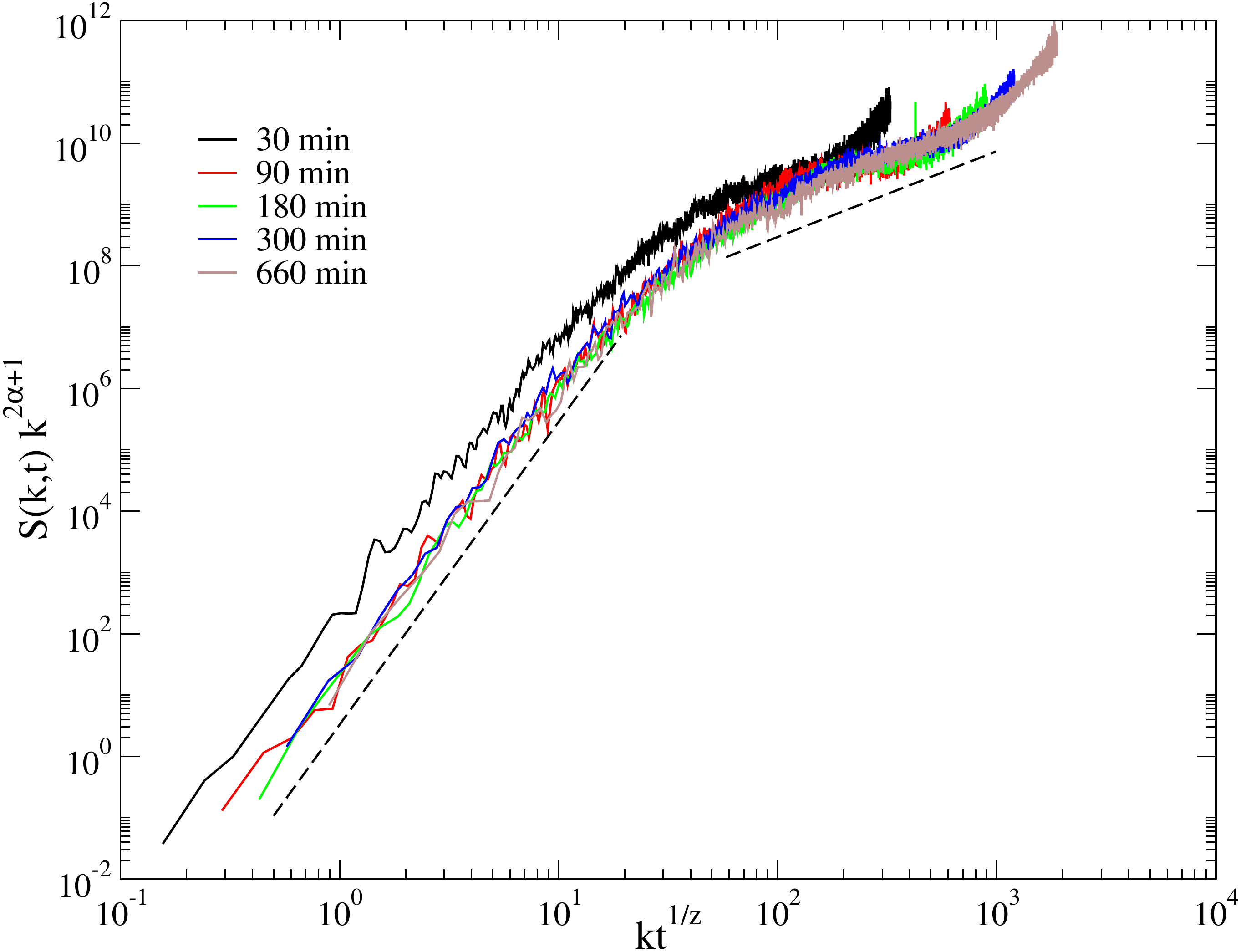}}
 \caption{(a) Power spectra of scanned profiles for $T=300~^\circ$C. Growth times increase 
from bottom to top. The dashed line has a slope -3.6 and represents the scaling 
$S\sim k^{-(2\alpha_s+1)}$.  
(b) Collapses of the power spectra using exponents $\alpha=1.94$ and $z=1.75$. 
Dashed lines represent the asymptotic behaviours of the scaling function $\Psi(x)$. 
The slopes are $2\alpha+1\approx 4.9$ and $2(\alpha-\alpha_s)\approx 1.3$.}
\end{figure}
 
\begin{figure}[ht]
 \centering
 \includegraphics[width=4.0cm]{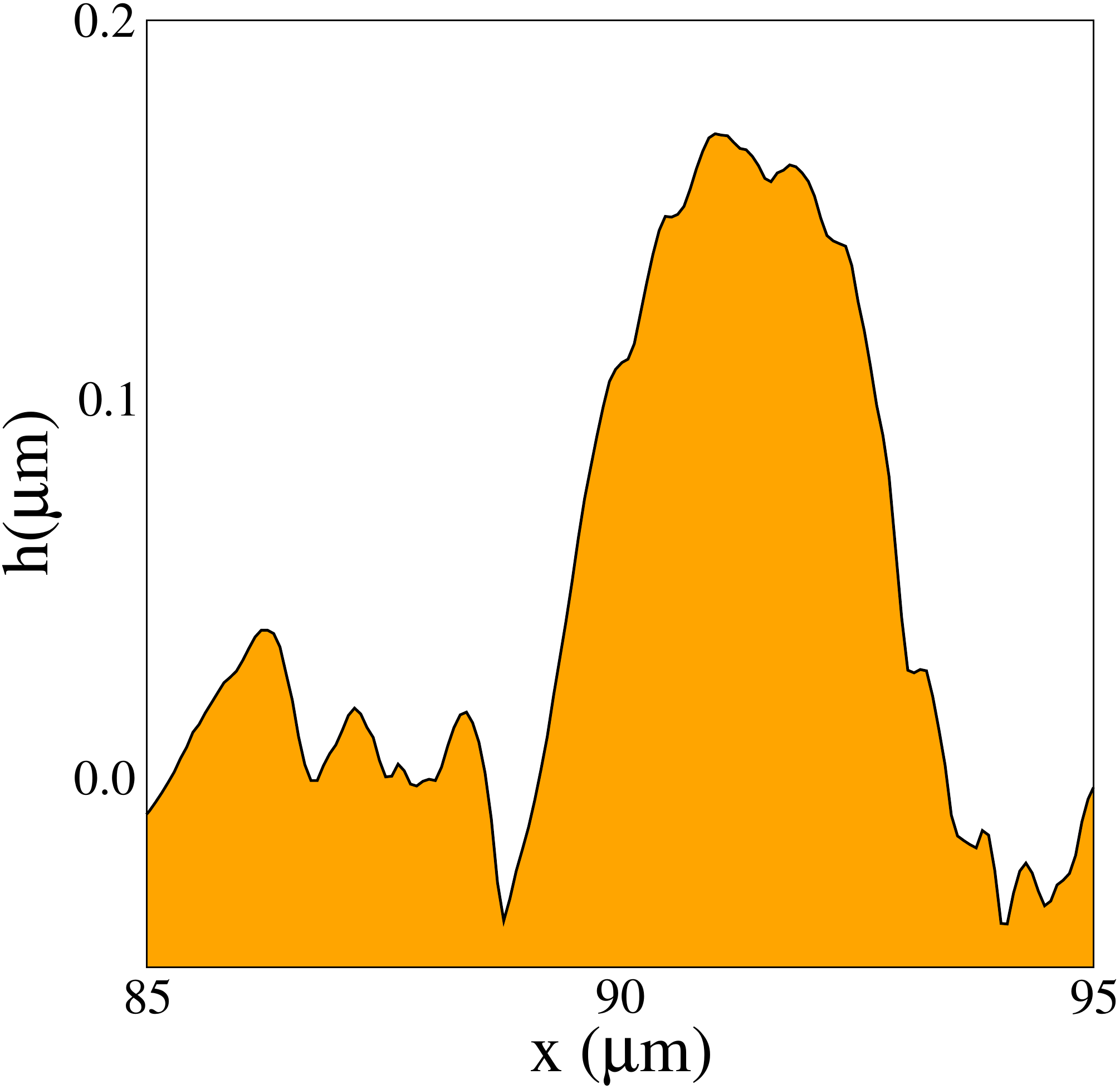}~
 \includegraphics[width=4.0cm]{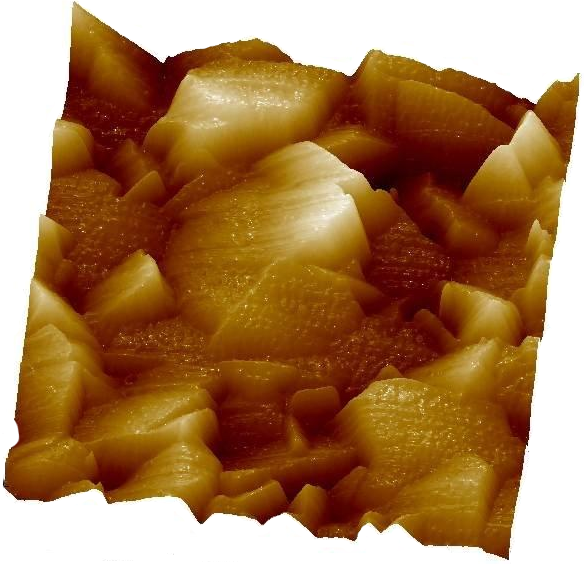} \\
\begin{small} (a)~~~~~~~~~~~~~~~~~~~~~~~~~~~~~~~(b)      \end{small}
 \caption{Surface morphologies for CdTe after a growth time of 300 min at a temperature of 300 $^\circ$C. (a) A 10 $\mu$m scan with using a profiler. (b)  A $3~\mu$m$~\times~3~\mu$m AFM image illustrating the faceted CdTe surface.}
 \label{fig:afm}
\end{figure}

Power spectra for $T=300~^\circ$C are shown in Fig.~\ref{fig:power300}. The curves exhibit 
the usual plateaus for small and power law  decays for large momenta. 
The power law yields $\alpha_s = 1.27(5)$ and, consequently, GDST 
states $\alpha_{loc}=1$. Notice that momenta in the range $k<2\pi/l_c$
were used in this analysis. Moreover, the curves shift upwards for large $k$ as time increases, 
representing $\alpha>\alpha_s$ since 
$S(k,t)\sim k^{2\alpha_s+1}t^{2(\alpha-\alpha_s)/z}$ in the scaling 
ansatz given by Eq. (\ref{eq:ansatz_power}). GDST thus predicts  surfaces with 
anomalous scaling and facets. 
Even though the profiler provides coarse-grained 
scans that hinder to reliably resolve faceted morphologies, as can be seen in Fig.~\ref{fig:afm}(a), 
a  AFM image with a scan step size of 5.8 nm and probe radius of 10~nm and, therefore, a resolution
about 100~times better than the profiler, clearly exhibits the faceted 
surface morphology predicted by GDST, as shown in Fig~\ref{fig:afm}(b). 
That is a remarkable feat of the GDST which 
has foreseen the faceted morphology in coarse-grained data with a poor (if any) resolution 
of the facets. From a theoretical point of view, this result corroborates that GDST connects 
the scaling properties of long wavelength fluctuations with local morphological properties. 
In our case,  the scaling properties of mesoscopic ($2\pi/l_c<k\ll2\pi/L$)
and macroscopic ($k\ll2\pi/l_c$) scales are connected with the surface morphology in a microscopic scale $k>2\pi/l_c$.  

\begin{figure}[ht]
 \centering
 \subfigure[\label{fig:corgrad}]{\includegraphics[width=7.5cm]{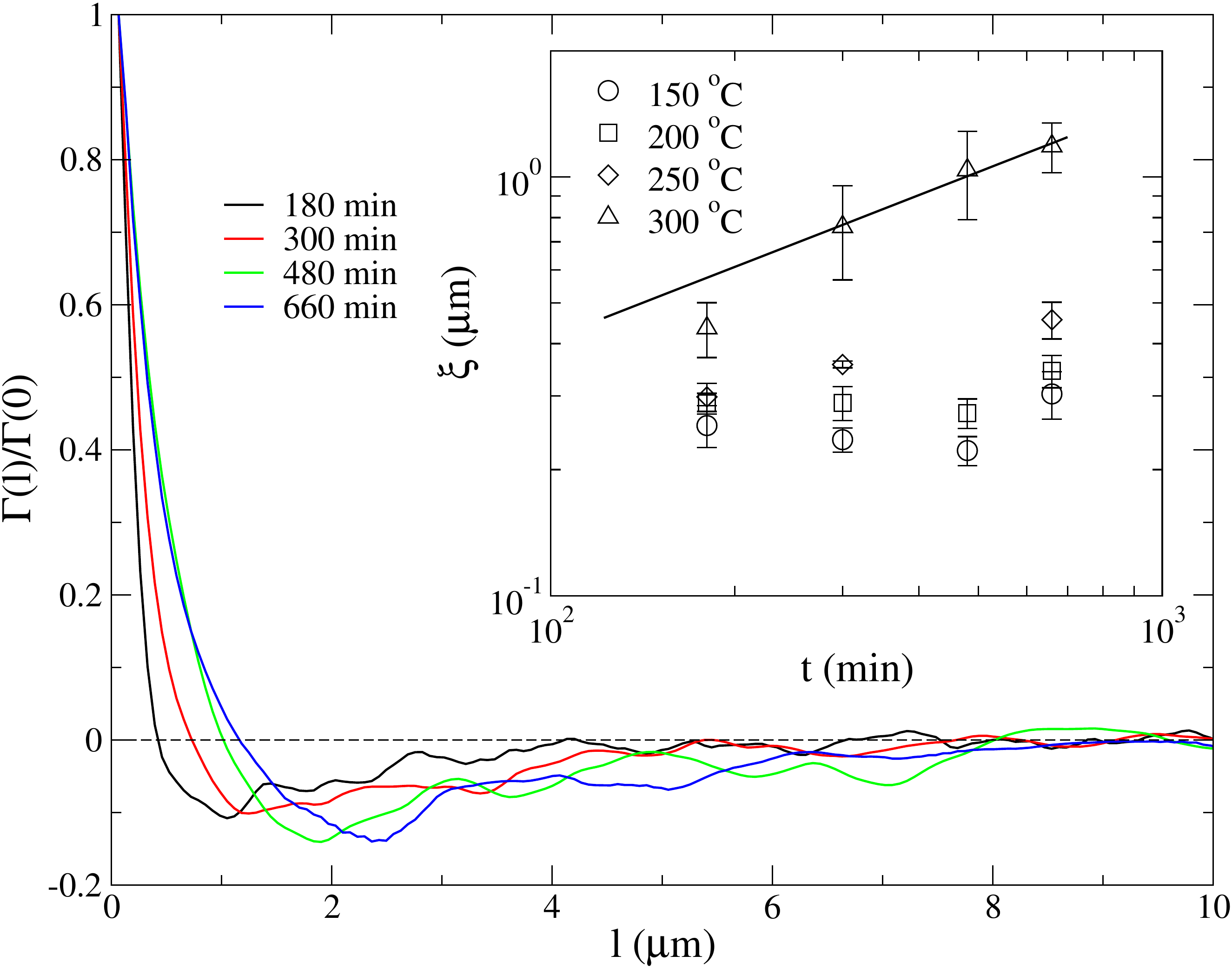}}
 \subfigure[ \label{fig:correlg}]{\includegraphics[width=7.5cm]{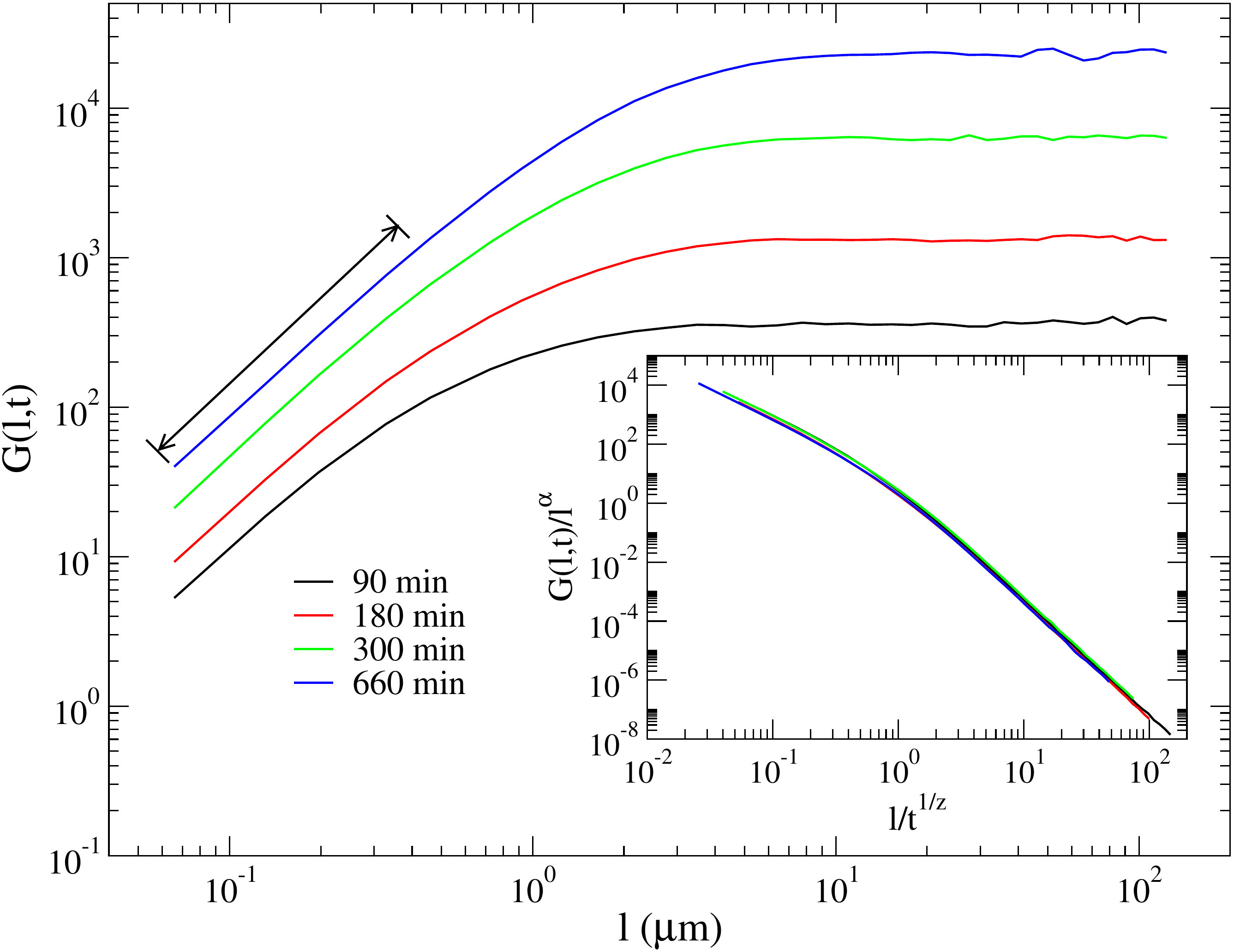}}
 \caption{(a) Slope-slope correlation functions for the CdTe surfaces. 
In the main plot, the correlation for distinct times and $T=300~^\circ$C are shown. Inset shows
 the correlation length against time. Straight line is a power law regression. (b) Height-height
 correlation functions for $T=300^\circ$ C. The interval of regression is indicated. Inset shows
 the collapse of the correlation functions using $\alpha=1.94$ and $z=1.75$}
\end{figure}

In Fig.~\ref{fig:colapses} we probe the scaling ansatz for $\Psi(x)$, given by 
Eq.~(\ref{eq:ansatz_power}), by plotting 
$k^{2\alpha+1}S(k,t)$  versus $k t^{1/z}$, $\alpha=\beta z$, where $z=1.75$  
was obtained via gradient correlation 
function [ explained in the next paragraph and shown in Fig.~\ref{fig:corgrad}] 
and $\beta=1.11$ via global interface  width (Figure~\ref{fig:widths}). 
The excellent collapse obtained for $t \ge 90$~min provides an additional evidence that the data is described by 
anomalous scaling for faceted surfaces. Notice that the curve corresponding to $t=30$~min, 
which was removed from the determination of the scaling exponents (Fig.~\ref{fig:widths}), 
does not collapse. Finally, the asymptotic scaling forms of $\Psi(x)$ are also 
confirmed as indicated by the dashed lines. 

The dynamic exponent $z$ can be determined using the slope-slope correlation function 
$\Gamma(l,t)= \langle \nabla h(x+l) \nabla h(x)\rangle$.
The correlation length $\xi$, which is assumed to scale as $\xi\sim t^{1/z}$ in GDST, can be defined as 
the first zero of the correlation function. The correlation functions and lengths are shown in Fig.~\ref{fig:corgrad}. There is no clear power law regime with $\xi>l_c$ for $T=150-250~^\circ$C. For  $300~^\circ$C, the slope provides $1/z = 0.57$.

Power spectrum analysis is itself sufficient to carry out the GDST, but correlation 
functions are widely used in the experimental investigations. Fig.~\ref{fig:correlg} shows 
that the height-height correlation function, equation~(\ref{eq:g}),  
also follows the usual qualitative behaviour of anomalous scaling, in which 
the curves are shifted upwardly as time evolves. 
However, a power law regression in the linear interval 
indicated turns out a self-affine surface with $\alpha_{loc} = 0.82(2)$, 
in disagreement with GDST. The scaling exponents for the other 
temperatures are shown in Table~\ref{tab:exp}.
The underestimated values of $\alpha_{loc}$ are reflecting the coarse-grained resolution 
of the scanning device since, in this scale, the analysed data are the convolution of 
surface and probe tip morphologies. 
The inset of Fig. \ref{fig:correlg} shows the correlation function collapses using the same
exponents that collapsed the power spectra, as shown in Fig.~\ref{fig:colapses}. The collapses are so or more convincing than those obtained for power spectra.

\begin{figure}[ht]
 \centering
 \subfigure[\label{fig:power250}]{\includegraphics[width=7.5cm]{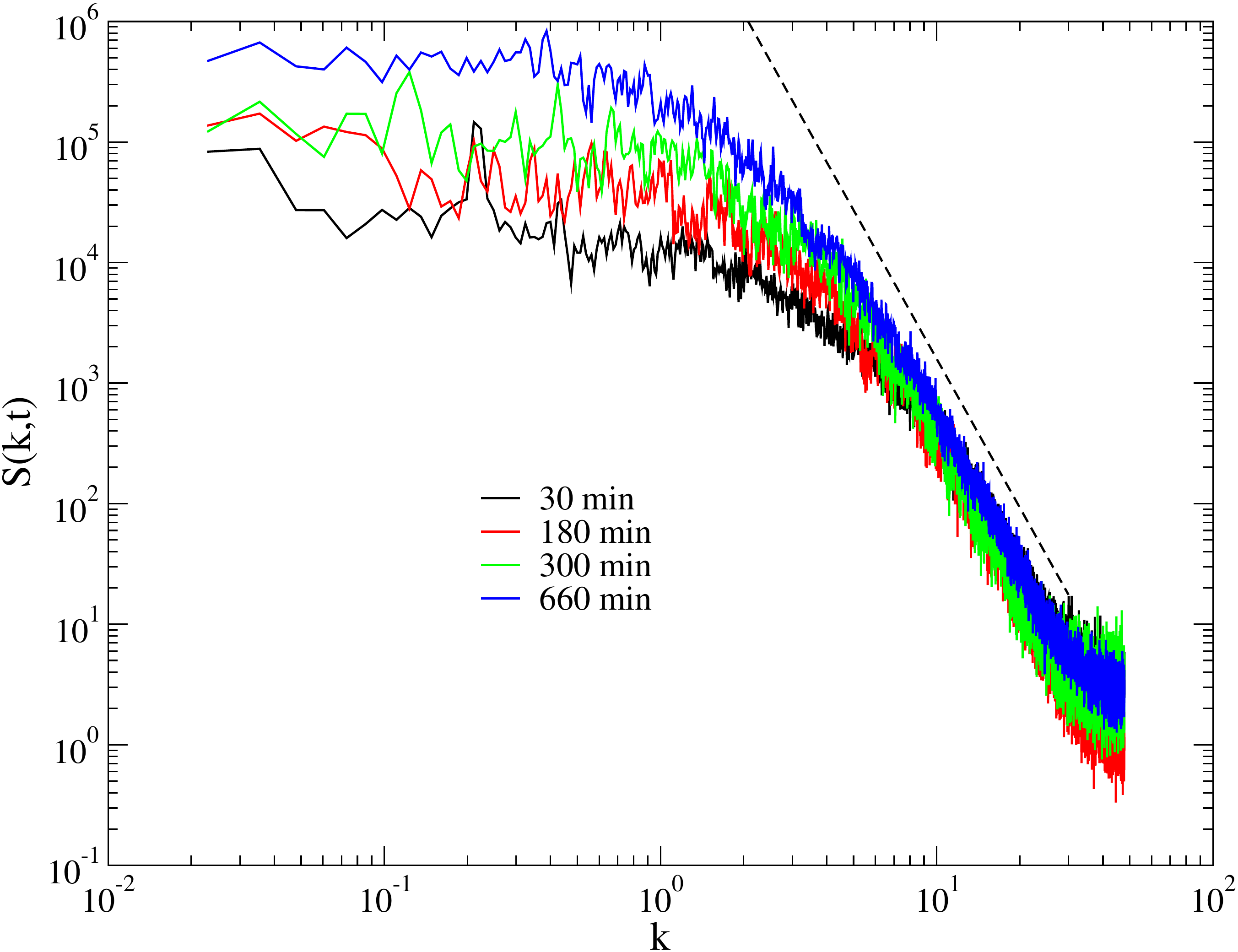}}
 \subfigure[ \label{fig:colapse250}]{\includegraphics[width=7.5cm]{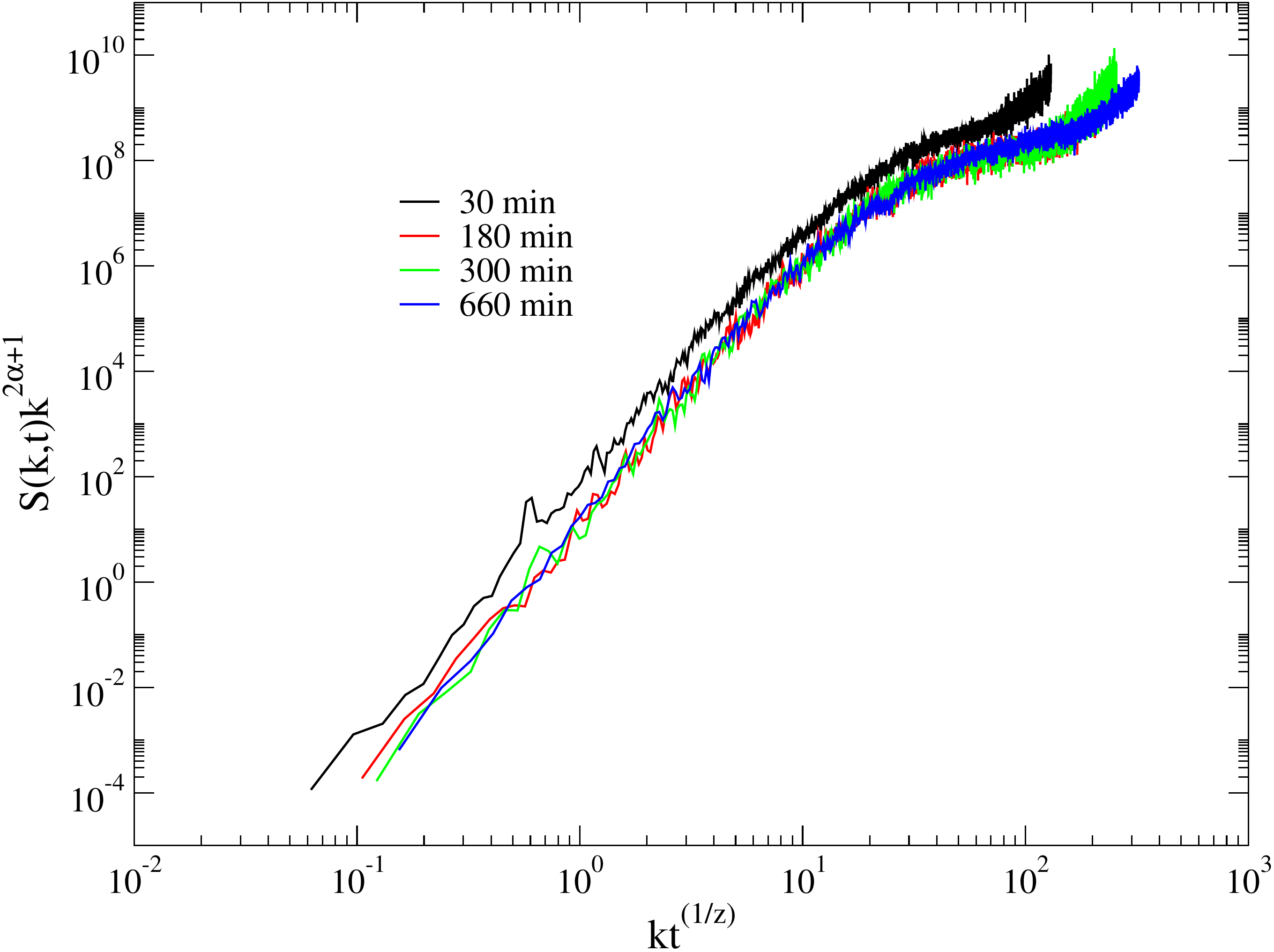}}
\caption{(a) Power spectra of scanned profiles for $T=250~^\circ$C. Growth times increase 
from bottom to top. The dashed line has a slope -4.1 and represents the scaling 
$S\sim k^{-(2\alpha_s+1)}$.
(b) Collapses of the power spectra using exponents $\alpha=2.2$ and $z=3.4$. }
\end{figure}

The power spectra for $T = 250~^\circ$C are shown in Fig.~\ref{fig:power250}. 
The curve for $t=600$~min  also exhibits a power decay in a range within $k<2\pi/l_c$ 
that yields $\alpha_s=1.6(1)$. Again, $S(k,t)$ shifts upwardly as time increases implying
$\alpha>\alpha_s$. Gradient-gradient correlation function does not  provide a reliable 
estimate of the the dynamical exponent $z$ [Fig.~\ref{fig:corgrad}] and thus 
a criterion of best collapse for $t>90$~min 
was used instead to obtain $z = 3.4(4)$ and $\alpha = 2.2(2)$. 
The collapse using these 
exponents are shown in Fig.~\ref{fig:colapse250}. The power spectrum analyses 
for $T=150$ and $200$~$^\circ$C did not provide reliable scaling regimes within 
the interval $k<2\pi/l_c$.
Even though the scaling for $T = 250~^\circ$C is less precise than for 
$T = 300~^\circ$C, it still points out a faceted anomalous scaling which was also
confirmed in AFM images.

The power spectra for $T=150$ and 200~$^\circ$C do not exhibit scaling properties
accurate enough to determine the exponents but the qualitative behaviour of faceted 
anomalous scaling, $\alpha_s>1$ and $\alpha>\alpha_s$, were also verified. AFM images
again confirms the faceted morphologies for these temperatures. Indeed, the effect of
temperature is to increase the size of faceted grains. For this reason, the scaling analysis
with profiler date is more accurate for higher temperatures.

\begin{table}[ht]
\caption{Scaling exponents for different growth temperatures. 
The numbers in parenthesis represent uncertainties in the last digit. 
The $\alpha_{loc}$ exponents were obtained using the height-height 
correlation function given by Eq. (\ref{eq:g}). 
The missing exponents are due to the absence of a reliable scaling regimes.}
\label{tab:exp}
\begin{center}
\begin{tabular}{ccccc}
\hline\hline
T ($^\circ$C) & $\beta$ & $\alpha_s$ & $z$ &  $\alpha_{loc}$ \\ \hline
150 & $0.59(9)$  & -- & -- &  $0.73(2)$\\
200 & $0.56(9)$  & -- & -- &  $0.76(3)$\\
250 & $0.65(2)$  & $1.6(1)$ & 3.4(4) & $0.77(2)$\\
300 & $1.11(2)$  & $1.27(5)$ & 1.75(5) & $0.82(2)$\\ \hline\hline
 \end{tabular}
\end{center}
\end{table}

\begin{figure}[ht]
 \centering
 \includegraphics[width=7.5cm]{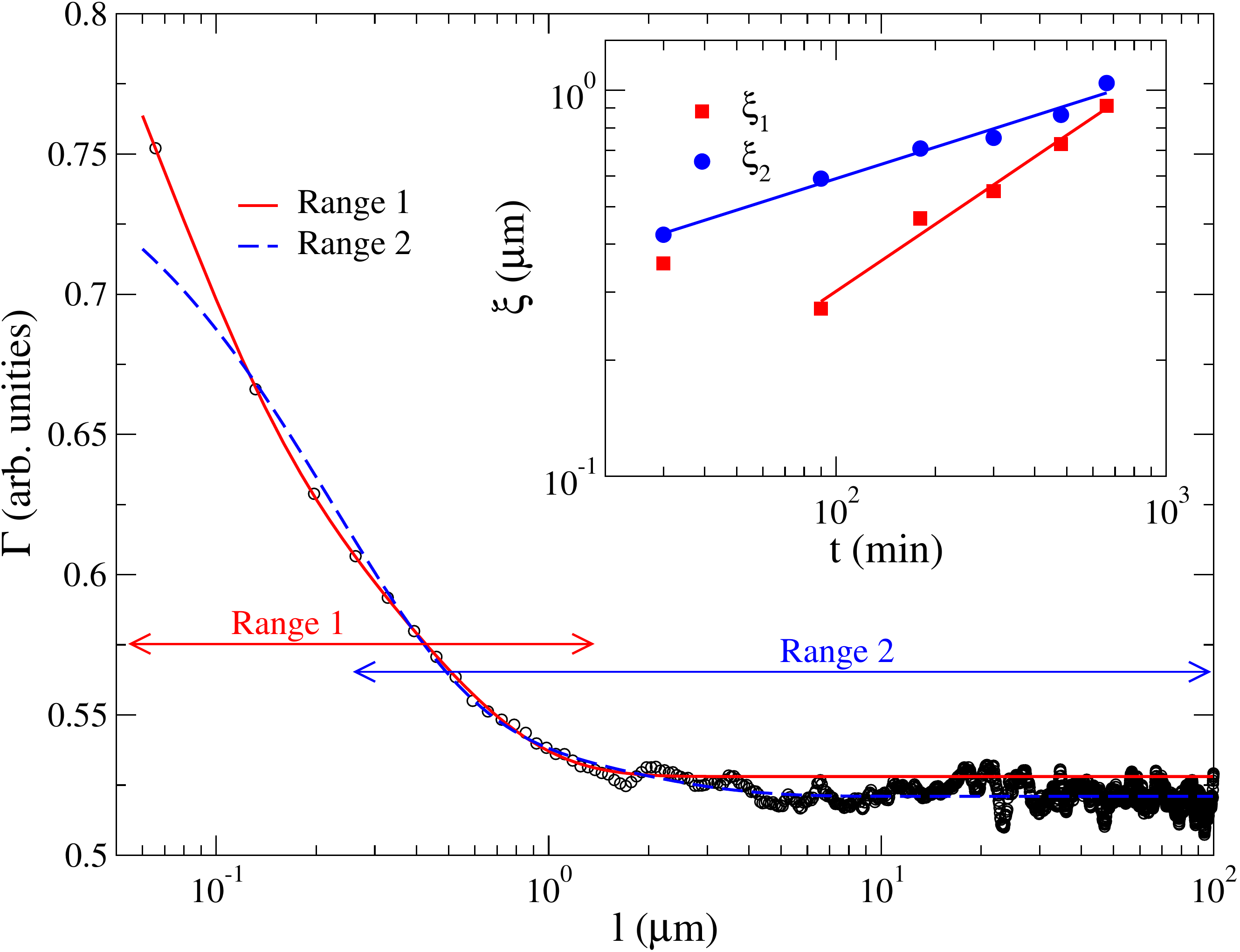}~
 \caption{Height-height correlation function of the CdTe surfaces. The growth time and temperature were $90$ min and $300^\circ$C. Circles represent the experimental data while solid lines the two-exponential regressions in the indicated intervals. Inset shows the correlation length obtained for the different regression intervals.}
 \label{fig:correl}
\end{figure}

Now, we compare the the GDST in the momentum space with the results 
for real space of this CdTe-TCO-glass system reported in Ref.~\cite{Mata2008}.
The scaling exponents $\alpha$ and $z$ shown in Table~\ref{tab:exp} are in disagreement 
with those in Ref.~\cite{Mata2008}, in which two sources of errors are present in the
exponent determination. 
The first one is the underestimation of the growth exponent used to determine 
$\alpha$ via scaling relation $\alpha=\beta z$. 
The second one is the correlation length $\xi$ that overestimated the dynamical exponent $z$,
as we clarify in this paragraph.
The correlation length was defined in Ref.~\cite{Mata2008} as the characteristic 
length of the decay in a two-point correlation function
$\Gamma(l,t) = \Pr[|h(x+l)-h(x)|\le m]$, where $m=0.1|h_{max} - h_{min}|$ 
and $\Pr[A]$ is the probability that the condition $A$ is satisfied. The usual correlation 
function $\Gamma=\langle h(x+l)h(x)\rangle$ undergoes exactly the same effects.
The correlation length was finally obtained by solving 
$\int_0^\xi \Gamma dl = 0.1\int_0^\infty\Gamma dl$, where $\Gamma(l)$ 
is a two-exponential fit to the experimental data.
However, if the tail corresponding to the long wavelength height fluctuations is or not 
left out of the regression, the exponents may change considerably. 
In Fig.~\ref{fig:correl}, we compare the regressions of the correlation functions in two ranges: discarding (range 1) and including (range 2) the tail.  The regression in the range 2, the same 
used in Ref.~\cite{Mata2008}, misfits the data only for $l \lesssim 0.3~\mu$m whereas 
the regression in the range 1 fits very well the small scales but deviates for 
$l\gtrsim2~\mu$m, which was the reason for the choice of range 2 in Ref. \cite{Mata2008}. 
The insertion to Fig.~\ref{fig:correl} shows the correlation 
lengths against time and the respective power law regressions. Both cases yield 
quite satisfactory scaling laws, $\xi_1\sim t^{0.58}$ and $\xi_{2}\sim t^{0.27}$ 
corresponding to ranges 1 and 2, respectively.
Assuming $\xi_i\sim t^{1/z_i}$, where $i=1,2$, we found $z_1=1.72$ and $z_2=3.69$. 
Notice that the regressions provided two characteristic lengths of the same magnitude ($1<\xi_2/\xi_1\lesssim 2$) but exhibiting very different scalings with time. 
Repeating this analysis for lower temperatures ($T=150-250~^\circ$C) we found 
$\xi_1\lesssim 0.5~\mu$m and no clear power law could be identified as observed 
for the gradient correlation function. Therefore, range 1 is the fit interval 
that yields dynamical exponents consistent with GDST. 

The conclusions of Ref.~\cite{Mata2008} are thus partially  modified since we cannot apply GDST 
with those  exponents to infer about which dynamical scaling regime the 
system belongs to. The exponent $z$ presented in Ref.~\cite{Mata2008} was misleadingly associate to the the dynamical exponent of the GDST.
It would be better to refer to this exponent as a long wavelength coarsening exponent, 
which increases with temperature whereas the actual dynamical exponent  
decreases. These results also stress out the difficulties usually observed in the fitting 
procedure of experimental data and the problems which can arise from a misinterpretation 
of the parameters obtained.  In this scenario, GDST using momentum space appears as a robust
method since local scaling properties can be obtained from mesoscopic and macroscopic scales.

In conclusion, we presented the scaling analysis of CdTe polycrystalline surfaces grown in glass
substrates using a generic dynamical 
scaling theory (GDST)~\cite{Ramasco}. We investigated both
height-height correlation functions and surface power spectra. 
The surfaces were scanned with a stylus profiler a resolution estimated as $l_c\approx 0.3~\mu$m. 

Surface power spectra exhibit reliable
scaling properties for $l>l_c$ at temperatures $250$ and $300~^\circ$C,
with a spectral roughness exponent $\alpha_s$ greater than 1 but different from the 
global roughness exponent $\alpha$. The GDST foresees therefore faceted morphologies with
local roughness exponent $\alpha_{loc}=1$ that represents locally smooth surfaces.
The height-height correlation function exhibits scaling only for $l\lesssim l_c$ and the
local roughness exponents are in the interval 0.7-0.8, indicating self-affine surfaces 
as described in a previous study of this system~\cite{Ferreira2006}.
However, high resolution AFM images confirm that the surfaces are faceted. 
The underestimated values of $\alpha_{loc}$ are reflecting the low resolution of the 
scanning device since, in this range, the analyzed profile is a convolution of
the surface and probe tip.

It is worth to note that  in a previous reports on anomalous scaling in faceted 
morphologies, investigated in the dissolution of polycrystalline iron~\cite{Cordoba}, 
the faceted anomalous scaling was evident only for one-dimensional 
sections in a direction orthogonal to the anisotropy. 
Differently, our scaling analysis of polycrystalline
CdTe consists in averaging over several randomly selected directions.

A renormalization group analysis of stochastic equations shows that the anomalous 
scaling cannot be present in local growth models \cite{Lopez2005}. Therefore, 
this theoretical result implies that disorder
and/or nonlocal effects are responsible by the anomalous scaling in experimental
systems. Indeed, the CdTe/TCO/Glass system undergoes both effects simultaneously. 
The amorphous glass 
substrate results in random growth orientations and, consequently, a polycrystalline
film is obtained. Moreover, the glass substrate and CdTe films have distinct 
coefficients of thermal expansion implying strained film. The interplay between
nonlocal strain and substrate disorder is a possible mechanism involved
in the anomalous scaling.

Our work opens new paths to investigate anomalous scaling experimentally. A natural extension
of this work, is to check the effect of the substrate on the scaling properties. Scaling analysis 
of AFM images may also complement the present investigation, in particular at lower temperatures 
where the scaling properties did  not allow to determine the scaling accurately.

\acknowledgments
This work was supported by the Brazilian agencies CNPq, FAPEMIG and CAPES. SCF thanks the kind hospitality at the Departament de F\'{\i}sica i Enginyeria Nuclear/UPC.


\end{document}